\documentclass[final,1p,times]{elsarticle}

\usepackage{amsmath}
\usepackage{amssymb}
\usepackage{epsfig}

\newcommand* {\bra}[1]{\ensuremath{\langle {#1} |}}
\newcommand* {\ket}[1]{\ensuremath{| {#1} \rangle}}

\journal{Physica B}

\begin{document}

\begin{frontmatter}

\title{Double detected spin-dependent quantum dot}

\author{J\'ozsef Zsolt Bern\'ad}

\ead{Zsolt.Bernad@physik.tu-darmstadt.de}

\address{Institut f\"{u}r Angewandte Physik, Technische Universit\"{a}t Darmstadt,
D-64289, Germany}
\address{Institute of Fundamental Sciences and MacDiarmid Institute for Advanced
Materials and Nanotechnology, Massey University, Private Bag 11~222, Palmerston
North 4442, New Zealand}

\begin{abstract}
We study the dynamics of a spin-dependent quantum dot system, where an unsharp and a sharp detection 
scenario is introduced. The back-action of the unsharp detection related to the magnetization, proposed in terms of the continuous 
quantum measurement theory, is observed via the von Neumann measurement (sharp detection) of the electric charge current. 
The behavior of the average electron charge current is studied as a function of the unsharp detection strength $\gamma$, and features 
of measurement back-action are discussed. The achieved equations reproduce the quantum Zeno effect. Considering magnetic leads, 
we demonstrate that the measurement process may freeze the system in 
its initial state. We show that the continuous observation may enhance the transition between spin states, in 
contradiction with rapidly repeated projective observations, when it slows down. Experimental
issue, such as the accuracy of the electric current measurement, is analyzed. 
\end{abstract}

\begin{keyword}
Quantum measurement theory \sep Quantum dot \sep Quantum Zeno effect \sep Heisenberg uncertainty principle

\end{keyword}

\end{frontmatter}

\section{Introduction}
The act of a quantum measurement is always performed by an external apparatus and involves complicated interactions with it. 
We are going to discuss the the back-action induced by the observer in the framework of unsharp measurements. The unsharp measurement 
extracts only partial information from  an observable, so we introduce another detector with sharp detection. We are interested in the 
output of the sharp detection, which means that the unsharp measurement will be treated in a \textit{nonselective} picture. 
The \textit{nonselective} description represents a measurement, where our record of data was lost and replaced by an average over the 
data ensemble. The sharp detector also 
has its back-action, projecting the system randomly into its eigenstate, but now the readout will be kept and the further evolution 
of the system will be neglected. The double detected setup gives a possibility to analyze the so-called quantum Zeno effect 
(QZE)~\cite{misra}. Rapidly repeated measurements give rise to the QZE, the suppression of transition 
between quantum states. In reality there are more complicated physical processes that take place during a quantum measurement, which can 
also cause QZE. The effect can be best understood in terms of the dynamic time evolution of the measured quantum systems.  \\
~\indent
The double detection scenario is similar to the ``indirect measurement'' process \cite{Mandelstam}, where the back action of a detector on 
the quantum system is observed by a third party, namely in our case by a sharp detector. In this context, 
the idea of detecting the measurement back-action related to one type of degree of freedom, with a detection scenario of an another type of 
degree of freedom, would be suitable for a future experiment. The charge detector is a convenient sharp detector type and its 
applicability in the semiconductor physics is very high. We choose the other detector to be a spin related, magnetization, detector. 
The application we have in mind is the spin-dependent single quantum dot, available in high quality due to 
massive progress in experimental technology. Spin manipulation and magnetization detection in quantum dot was studied in experiment by 
Ref. \cite{Experiment}. An external field, used for spin manipulation, can be viewed as an environment of the subsystem, the quantum dot. 
The whole of quantum system dynamics is reversible. Tracing out the environment's degrees of freedom, we arrive at a non-unitary 
time evolution \cite{BreuerPetruccione}. In general, all these non-unitary processes are connected through the Kraus-form (\ref{GM}),
 related to the completely positive mappings of the density matrices \cite{Kraus}. The subsystem's non-unitary dynamic, imposed by the 
external field, can be interpreted as an unsharp measurement \cite{nuclearmagnetic}. These manipulations are time-continuous, so we will 
study the magnetization detection in the frame of the continuous quantum measurement theory~\cite{cont.model}, 
avoiding the modeling of the detector system as a quantum system.\\
~\indent 
While the system described above is similar to the spin-to-charge conversion in quantum dots~\cite{Elzerman, Barrett}, the model of 
the unsharp detector is different. Here, the back-action of continuous quantum measurement on magnetization is 
investigated by a detected electric current, focusing on measurement back-action and on QZE in the spin states. The unsharp detector 
of the spin-to-charge 
conversion is modeled as a quantum point contact with a fixed coupling Hamiltonian. The theory presented here is 
broader, because the coupling between the unsharp detector and the system is choice of will, however has to be subject to a real 
experimental setup.\\
~\indent
For a possible experimental setup, the idea of the double detection in a spin-dependent quantum dot is reasonable, because quantum 
dots and also spin manipulations are important for the realization of qubits. The effects of spin decoherence related to quantum 
computing was studied by Refs. \cite{Kane,Loss,Jelezko}. On the other hand the field of indirect measurements on quantum dots by 
means of Coulomb-coupled, quantum point contacts, single-electron transistors, or double quantum dot's \cite{Barrett,Gur97}-\cite{Mao04} 
and the effects of QZE~\cite{Gur97, QZE} was studied by several works, and the concept of time-continuous measurement has been part of 
this field \cite{Barrett,Gur97,Kor99,Goa01,BBDG}.  Previous research has monitored the sharp or unsharp detection of the observables 
related to the electron charge. While the work presented here examines the sharp detection of the electric current, it also contains 
an unsharp detection of a spin observable, the magnetization. The model proposed here is similar to the work of Ref. \cite{BBDG}, 
where the authors studied the effect of the unsharp detection of electric current in a \textit{selective} measurement scenario. For the difference from 
that work, we applied the model for the magnetization detection and we studied the evolution of the density matrix as a function of
the electron number tunneled through the system.\\
~\indent
This article is organized as follows. In Sec. \ref{The model} we introduce our model. We derive the many-body Scr\"odinger equation including
the terms of the continuous quantum measurement. We represent the density matrix as a function of the electron number in the right lead. 
The results of measurement back-action are shown and discussed in Secs. \ref{Results}, \ref{Conclusions}. We investigate the accuracy of 
the electric charge current measurement in Sec. \ref{accuracy}. General and continuous 
quantum measurement theories have a wide literature and to ensure the background of this work we present a short summary of this topic 
in the \ref{GM} and \ref{CQM} using Refs. \cite{cont.model, diosik, NielsenChuang}.

\section{The model}
\label{The model}
\begin{figure}[t]
\begin{center}
\includegraphics[width=2.51in]{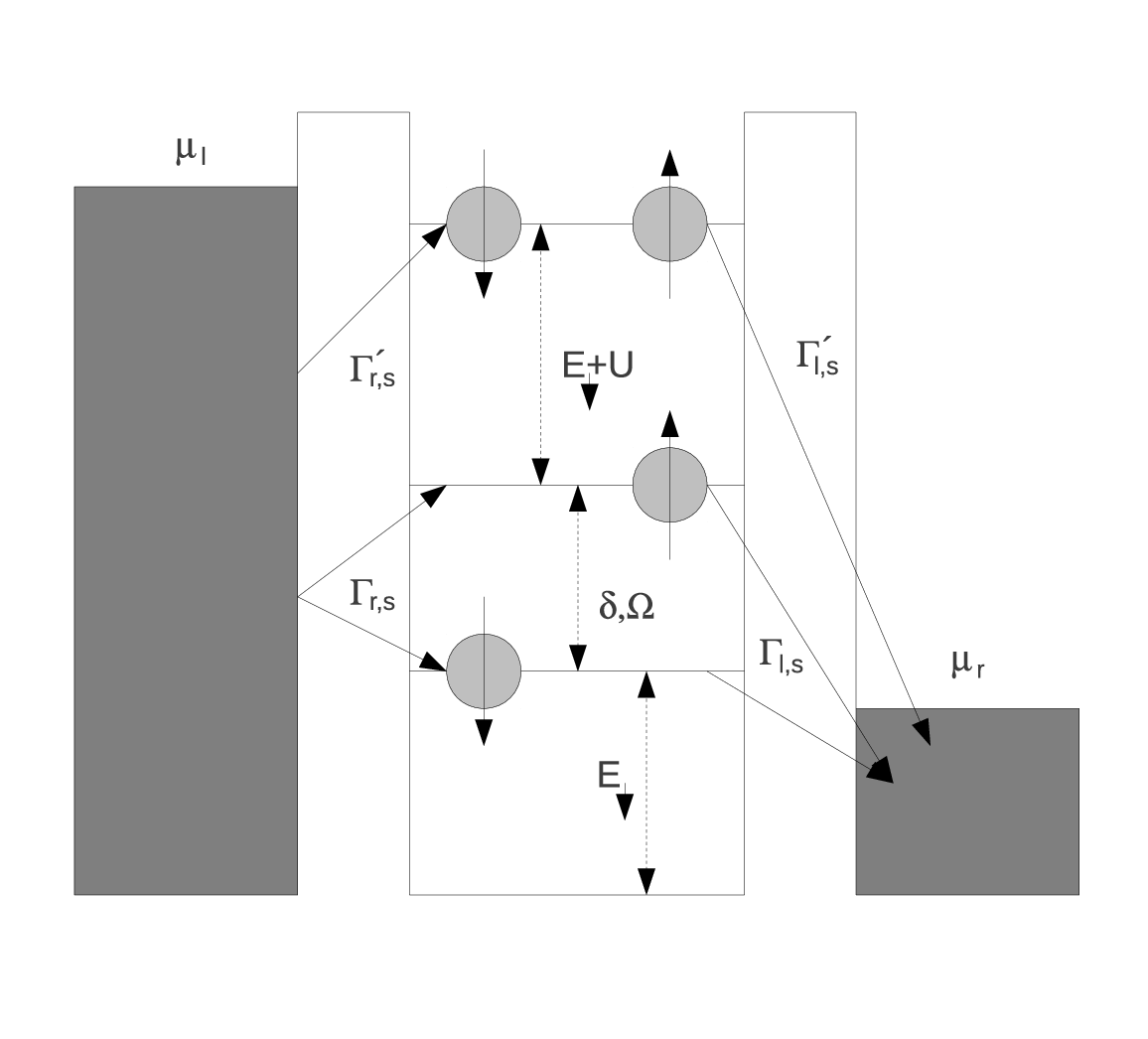}
\caption{\label{Schematic}Schematic of the spin-dependent quantum dot model, including the left-to-right tunneling assumption and the energy differences
between the considered levels.} 
\end{center}
\end{figure}

We consider the spin-dependent quantum dot, subject of experimental work \cite{KoppensScience309,PettaScience309}, which is coupled to 
two separate electron reservoirs. The density of states in the reservoirs is very high (continuum), and the dot contains only isolated 
levels. 
We consider the highest energy level to be the state of two electrons with different spins ($s=\pm 1/2$) and therefore we include the effects of 
Coulomb interaction. The split of the one-electron energy level is done by a $z$-directed magnetic field $B_z$. 
We include a $x$-directed magnetic field $B_x$, describing the coherent oscillations between the spin up and spin down levels. 
The full Hamiltonian of the system reads
\begin{equation}
\label{Hamiltonian}
\hat{H}=\hat{H}_D+\hat{H}_R+\hat{H}_I, 
\end{equation}
where
\begin{eqnarray}
\hat{H}_D&=&\sum_{s}E_s \hat{a}^{\dagger}_{D,s}\hat{a}_{D,s}+
U\hat{a}^{\dagger}_{D,\frac{1}{2}}\hat{a}_{D,\frac{1}{2}}\hat{a}^{\dagger}_{D,-\frac{1}{2}}\hat{a}_{D,-\frac{1}{2}} \nonumber \\
&+& \hbar \Omega (\hat{a}^{\dagger}_{D,\frac{1}{2}}\hat{a}_{D,-\frac{1}{2}}+\hat{a}^{\dagger}_{D,-\frac{1}{2}}\hat{a}_{D,\frac{1}{2}})  
\end{eqnarray}
is the Hamiltonian of the quantum dot, 
\begin{equation}
\hat{H}_R=\sum_{l,s} E_{l,s}\hat{a}^{\dagger}_{l,s}\hat{a}_{l,s}+\sum_{r,s} E_{r,s}\hat{a}^{\dagger}_{r,s}\hat{a}_{r,s}  
\end{equation}
is the Hamiltonian of the reservoirs (leads), and 
\begin{eqnarray}
\hat{H}_I&=& \sum_{l,s} \hbar(\omega_{l,s} \hat{a}^{\dagger}_{l,s}\hat{a}_{D,s} + \omega^*_{l,s}
              \hat{a}^{\dagger}_{D,s}\hat{a}_{l,s}) \nonumber \\
              &+& \sum_{r,s} \hbar (\omega_{r,s}\hat{a}^{\dagger}_{r,s}\hat{a}_{D,s} + \omega^*_{r,s}
              \hat{a}^{\dagger}_{D,s}\hat{a}_{r,s})
\end{eqnarray}
is the coupling Hamiltonian between the reservoirs and dot. The subscripts $l$ and
$r$ enumerate correspondingly the (very dense) levels in the
left and right leads. $\hat{a}_{D,s}$($\hat{a}^{\dagger}_{D,s}$) is the annihilation (creation) operator
of spin $s$ in the quantum dot. $\hat{a}_{l/r,s}$($\hat{a}^{\dagger}_{l/r,s}$) is the annihilation (creation) operator
of spin $s$ in the reservoir $l$ or $r$. 
$U$ is the Coulomb repulsion energy,  the energy difference $E_{\frac{1}{2}}-E_{-\frac{1}{2}}$ is proportional to $B_z$ and the frequency 
$\Omega \sim B_x$.
$E_{l,s}$ and $E_{r,s}$ are the one-electron energies with spin $s$ in the left and right leads.
$\hbar \omega_{l,s}$ and $\hbar \omega_{r,s}$ are the respective tunneling 
amplitudes of spin $s$ between the left or right reservoir and the dot. (Fig. \ref{Schematic}) \\
For simplicity, we restrict ourselves to a low temperature case, $T \to 0$. All the levels in the right and left lead are initially filled with 
electrons up to the Fermi energy $\mu_r$ and $\mu_l$, respectively. This situation will be treated as the ``vacuum'' state $\ket{0}$. 
We consider a large bias and that the energy levels are inside the band, $\mu_l \gg E_s,U \gg \mu_r$. In the context of these conditions, 
the electric current flows only from left to right. The evolution of the whole system is described by the many-particle wave function. 
Taking into account the assumptions, the wave function is represented as 
\begin{eqnarray}
\ket{\Psi(t)} & = & \left [ b_0(t) + \sum_l b_{l,\frac{1}{2}}(t)
     a_{D,\frac{1}{2}}^{\dagger}a_{l,\frac{1}{2}} + \sum_{l,s;r,s} b_{lr,s}(t)a_{r,s}^{\dagger}a_{l,s}+\sum_l b_{l,-\frac{1}{2}}(t)
    a_{D,-\frac{1}{2}}^{\dagger}a_{l,-\frac{1}{2}} \right. \nonumber\\
     &+&\left. \sum_{l,-s<l',s} b_{ll',s}(t)a_{D,\frac{1}{2}}^{\dagger}a_{D,-\frac{1}{2}}^{\dagger}
     a_{l,-s}a_{l',s} 
      +  \sum_{l,-s<l',s;r} b_{ll'r,\frac{1}{2}}(t)
           a_{D,\frac{1}{2}}^{\dagger}a_{r,-\frac{1}{2}}^{\dagger}a_{l,-s}a_{l',s}  \right.  \label{Psi}\\ 
    &+& \left. \sum_{l,s<l',s';r,s>r',s'}b_{ll'rr',ss'}(t)\hat{a}^\dagger_{r,s}\hat{a}^{\dagger}_{r',s'}\hat{a}_{l,s}\hat{a}_{l',s'}
       +\sum_{l,-s<l',s;r} b_{ll'r,-\frac{1}{2}}(t)a_{D,-\frac{1}{2}}^{\dagger}a_{r,\frac{1}{2}}^{\dagger}a_{l,-s}a_{l',s}
           + \ldots \right ] |0\rangle, \nonumber 
\end{eqnarray}
where, for example $\sum_{l,-s<l',s}$ is the sum over all states with energy $E_{l,-s}$ and $E_{l',s}$ and with the condition $E_{l',s}>E_{l,-s}$. 
The amplitudes represent the physical situations as: $|b_0(t)|^2$ the probability that the system is in the ``vacuum'' state at time $t$, 
$| b_{l,\frac{1}{2}}(t)|^2$ the probability that one electron with spin up was annihilated in the left reservoir 
and one electron with spin up created in the quantum dot at time $t$, and so on. The
``vacuum'' state $\ket{0}$ in this representation has the following properties:
\begin{eqnarray}
 \hat{a}^\dagger_{l,s}\ket{0}&=&0,\,\,\hat{a}_{r,s}\ket{0}=0,\,\,\hat{a}^\dagger_{l,s}\hat{a}_{l,s}\ket{0}=\ket{0}, \nonumber \\
\hat{a}_{r,s}\hat{a}^\dagger_{r,s}\ket{0}&=&\ket{0},\,\,\hat{a}_{D,s}\hat{a}^\dagger_{D,s} \ket{0}=\ket{0},\,\,.\,.\,.\,.\,.
\end{eqnarray}
%%%%%%%%%%%%%%%%%%%%%%%%%%%%%%%%%%%%%%%%%%%%%%%%%%%%%%%%%%%%%%%%%%%%%%%%%%%%%%%
%%%%%%%%%%%%%%%%%%%%%%%%%%%%%%%%%%%%%%%%%%%%%%%%%%%%%%%%%%%%%%%%%%%%%%%%%%%%%%%%%%
~\indent
Now, we apply the theory of continuous quantum measurement (\ref{CQM}). The time evolution of the system in the presence 
of a time-continuous measurement of the magnetization, 
$\hat{M}=\hat{a}^{\dagger}_{D,\frac{1}{2}}\hat{a}_{D,\frac{1}{2}}-\hat{a}^{\dagger}_{D,-\frac{1}{2}}\hat{a}_{D,-\frac{1}{2}}$, is given by a modified Schr\"odinger equation:
\begin{eqnarray}
\label{Schrodinger}
d| \Psi \rangle=\Big[&&-\frac{i}{\hbar}\hat{H}dt-\frac{\gamma}{8}\left(\langle \hat{M}\rangle-\hat{M}
\right)^2dt  \nonumber \\
&&-\frac{\sqrt{\gamma}}{2}\left(\langle \hat{M}\rangle-\hat{M}
\right)dW\Big] | \Psi \rangle,
\end{eqnarray}
where $\langle \hat{M}\rangle=\langle \Psi| \hat{M}| \Psi \rangle$ is average detected magnetization and $W$ is the Wiener process. In order to derive this equation, we assumed that the  
detector bandwidth is bigger than the eigenfrequencies of the system, defined by the Hamiltonian $\hat{H}$.
The main parameter of the theory, the detection performance 
(or detection strength), is defined as
\begin{equation}
\label{gamma}
\gamma=\frac{1}{\Delta t(\Delta M)^{2}}, 
\end{equation}
where $\Delta t$ is the time-resolution (or, equivalently, the inverse bandwidth) of the detector (detecting the magnetization $\hat{M}$) and 
$\Delta M$ is the statistical error characterizing unsharp detection of the average value of the magnetization $\hat{M}$ 
in the period $\Delta t$.\\
Substituting eq.~(\ref{Psi}) into the equation of motion \eqref{Schrodinger} using the 
Hamiltonian (\ref{Hamiltonian}), we find a system of coupled differential equations for the amplitudes
 \begin{eqnarray}
d b_{0}(t) &=&-i \sum_l \omega_{l,\frac{1}{2}}{b}_{l,\frac{1}{2}}(t)dt-i\sum_l \omega_{l,-\frac{1}{2}}{b}_{l,-\frac{1}{2}}(t)dt-
\frac{\gamma}{8}\langle \hat{M}\rangle^2 b_{0}(t) dt- \frac{\sqrt{\gamma}}{2}\langle \hat{M}\rangle b_{0}(t) dW ,\\
d b_{l,\frac{1}{2}}(t)&=& -i \frac{E_{\frac{1}{2}}-E_l}{\hbar}{b}_{l,\frac{1}{2}}(t)dt-i\omega^*_{l,\frac{1}{2}} {b}_{0}(t)dt -i  
\Omega {b}_{l,-\frac{1}{2}}(t)dt -i \sum_{l'} \omega_{l',-\frac{1}{2}}{b}_{ll',-\frac{1}{2}}(t) dt \nonumber \\
&-& i\sum_{r} \omega^*_{r,\frac{1}{2}} b_{lr,\frac{1}{2}}(t)dt 
- \frac{\gamma}{8}\left(\langle \hat{M}\rangle-1\right)^2b_{l,\frac{1}{2}}(t)dt-\frac{\sqrt{\gamma}}{2}
\left(\langle \hat{M}\rangle-1\right) b_{l,\frac{1}{2}}(t)dW, \\
d b_{lr,s}(t)&=&-i\frac{E_{r,s}-E_{l,s}}{\hbar}{b}_{lr,s}(t)dt-i\omega_{r,s} {b}_{l,s}(t)dt-i\omega_{r,-s} {b}_{l,-s}(t)dt-i
\sum_{l'} \omega_{l',s}{b}_{ll'r,s}(t)dt \nonumber \\
&-&i\sum_{l'} \omega_{l',-s}{b}_{ll'r,-s}(t)dt 
-\frac{\gamma}{8}\langle \hat{M}\rangle^2 b_{lr,s}(t) dt- \frac{\sqrt{\gamma}}{2}\langle \hat{M}\rangle b_{lr,s}(t) dW ,\\
d b_{l,-\frac{1}{2}}(t)&=& -i \frac{E_{-\frac{1}{2}}-E_l}{\hbar}{b}_{l,-\frac{1}{2}}(t)dt-i\omega^*_{l,-\frac{1}{2}} {b}_{0}(t)dt -i  
\Omega {b}_{l,\frac{1}{2}}(t)dt -i \sum_{l'} \omega_{l',\frac{1}{2}}{b}_{ll',\frac{1}{2}}(t) dt \nonumber \\
&-& i\sum_{r,-\frac{1}{2}} \omega^*_{r} b_{lr,-\frac{1}{2}}(t)dt 
- \frac{\gamma}{8}\left(\langle \hat{M}\rangle+1\right)^2b_{l,-\frac{1}{2}}(t)dt-\frac{\sqrt{\gamma}}{2}
\left(\langle \hat{M}\rangle+1\right) b_{l,-\frac{1}{2}}(t)dW, 
\end{eqnarray}
 \begin{eqnarray}
d b_{ll',s}(t)&=&-i\frac{E_s+E_{-s}+U-E_l-E_{l'}}{\hbar}{b}_{ll',s}(t)dt-i\omega^*_{l',s} {b}_{l,s}(t)dt-i\omega^*_{l,s} {b}_{l',s}(t)dt
\nonumber \\
&-&i\omega^*_{l',-s} {b}_{l,-s}(t)dt-i\omega^*_{l,-s} {b}_{l',-s}(t)dt 
-i\sum_{r} \omega^*_{r,s}{b}_{ll'r,s}(t)dt-i\sum_{r} \omega^*_{r,-s}{b}_{ll'r,-s}(t)dt \nonumber \\
&-&\frac{\gamma}{8}\langle \hat{M}\rangle^2 b_{ll',s}(t) 
dt-\frac{\sqrt{\gamma}}{2}\langle \hat{M}\rangle b_{ll',s}(t) dW,\\
\cdots\cdots\cdots&\cdots&\cdots\cdots\cdots\,\,\,\cdots\,\,\,\cdots\cdots 
\nonumber
\end{eqnarray}
The sharp measurement is represented by a electric current measurement, which is related to the accumulated charge in the right lead. 
In order to analyze this quantity we introduce the density matrix as a function of $n$, the number of electrons in the 
right lead. The Fock space of the quantum
dot consists of only four possible states, namely:  $|a \,\rangle=|0\rangle$ the dot is empty, $|b \,\rangle=|\uparrow\rangle$ 
the dot contains a spin up electron ($s=\frac{1}{2}$), 
$|c\, \rangle=|\downarrow \rangle$ the dot contains a spin down electron($s=-\frac{1}{2}$) and 
$|d \,\rangle=| \uparrow \downarrow \rangle$ is the fully occupied dot. In our notation, these probabilities are represented as follows:
\begin{equation}
\sigma_{aa}=\sum_n\sigma_{aa}^{(n)} 
= |b_{0}(t)|^2 + \sum_{l,s;r,s} |b_{lr,s}(t)|^2  + \ldots
\end{equation}
\begin{eqnarray}
\sigma_{bb} &=&\sum_n\sigma_{bb}^{(n)}= \sum_l |b_{l,\frac{1}{2}}(t)|^2 +
             \sum_{l,-s<l',s;r} |b_{ll'r,\frac{1}{2}}(t)|^2  \nonumber \\ &+& \ldots
\end{eqnarray}
\begin{eqnarray} 
\sigma_{bc} &=&\sum_n\sigma_{bc}^{(n)}= \sum_l b_{l,\frac{1}{2}}(t)b_{l,-\frac{1}{2}}^*(t)  \nonumber \\
            &+&
             \sum_{l,-s<l',s;r} b_{ll'r,\frac{1}{2}}(t) b_{ll'r,-\frac{1}{2}}^*(t)+ \ldots
\end{eqnarray}
\begin{eqnarray} 
\sigma_{cc} &=&\sum_n\sigma_{cc}^{(n)}= \sum_l |b_{l,-\frac{1}{2}}(t)|^2 +
             \sum_{l,-s<l',s;r} |b_{ll'r,-\frac{1}{2}}(t)|^2  \nonumber \\ &+& \ldots
\end{eqnarray}
\begin{equation} 
\sigma_{dd} =\sum_n\sigma_{dd}^{(n)}= \sum_{l,-s<l',s} |b_{ll',s}(t)|^2 + \ldots 
\end{equation} 
We are going to investigate a nonselective continuous quantum measurement case, which means that we are only interested in the average over 
the different realization of the wave function $\ket{\Psi}$. \textit{As a first step}, we apply the quantum Ito rules~\cite{HudsonP} 
for the product rule of differentiation,
\begin{equation}
 d\big(\ket{\Psi}\bra{\Psi}\big)=d \ket{\Psi} \bra{\Psi}+ \ket{\Psi}d\bra{\Psi}+ d\ket{\Psi} d\bra{\Psi}. 
\end{equation}
Through this step the stochastic time evolution of the density matrix is defined. Now, we average over the realizations~\cite{BMZ}, where 
we use that $W$ is the standard Wiener process, a Gaussian random variable with zero mean value ($\mathbf{M}$) and variance $t$,
\begin{equation}
\mathbf{M}(dW)=0,\,\,d^2W=dt,\,\,d^nW=0, \,n>2. 
\end{equation}
In the context of the nonselective measurement we define the following matrix elements:
\begin{eqnarray}
\label{rho}
\rho_{aa}^{(n)}&=&\mathbf{M}(\sigma_{aa}^{(n)}),\,\rho_{bb}^{(n)}=\mathbf{M}(\sigma_{bb}^{(n)}),\,
\rho_{bc}^{(n)}=\mathbf{M}(\sigma_{bc}^{(n)}), \nonumber \\
\rho_{cc}^{(n)}&=&\mathbf{M}(\sigma_{cc}^{(n)}),\, \rho_{dd}^{(n)}=\mathbf{M}(\sigma_{dd}^{(n)}).
\end{eqnarray}
\textit{As a second step}, we use the large bias assumption \cite{Gur96} and a straightforward calculation yields a chain differential equations
for the density matrix elements defined in eq. \eqref{rho}
\begin{eqnarray}
\dot\rho_{aa}^{(n)} & = &-(\Gamma_{l,\frac{1}{2}}+\Gamma_{l,-\frac{1}{2}})\rho_{aa}^{(n)}
+\Gamma_{r,\frac{1}{2}}\rho_{bb}^{(n-1)}+\Gamma_{r,-\frac{1}{2}}\rho_{cc}^{(n-1)}+\sqrt{\Gamma_{r,\frac{1}{2}}\Gamma_{r,-\frac{1}{2}}}
(\rho_{bc}^{(n-1)}+\rho_{cb}^{(n-1)}),
\label{eqan}\\
\dot\rho_{bb}^{(n)} & = &i\Omega(\rho_{bc}^{(n)}-\rho_{cb}^{(n)})-(\Gamma'_{l,-\frac{1}{2}}+\Gamma_{r,\frac{1}{2}}) 
\rho_{bb}^{(n)}+\Gamma_{l,\frac{1}{2}}\rho_{aa}^{(n)}
+\Gamma'_{r,-\frac{1}{2}}\rho_{dd}^{(n-1)} \nonumber \\&-&\frac{\sqrt{\Gamma_{r,\frac{1}{2}}\Gamma_{r,-\frac{1}{2}}}+
\sqrt{\Gamma'_{l,\frac{1}{2}}\Gamma'_{l,-\frac{1}{2}}}}{2}(\rho_{bc}^{(n)}+\rho_{cb}^{(n)}),
\label{eqbn}\\
\dot\rho_{bc}^{(n)}&=&-i\delta \rho_{bc}^{(n)}+i\Omega(\rho_{bb}^{(n)}-\rho_{cc}^{(n)})+\sqrt{\Gamma_{l,\frac{1}{2}}\Gamma_{l,-\frac{1}{2}}}
\rho_{aa}^{(n)}-\frac{\Gamma'_{l,\frac{1}{2}}+\Gamma_{r,-\frac{1}{2}}}{2}\rho_{bb}^{(n)}-
\frac{\Gamma'_{l,-\frac{1}{2}}+\Gamma_{r,\frac{1}{2}}}{2}\rho_{cc}^{(n)} \nonumber \\
&-&\frac{\Gamma'_{l,\frac{1}{2}}+
\Gamma'_{l,-\frac{1}{2}}+\Gamma_{r,\frac{1}{2}}+\Gamma_{r,-\frac{1}{2}}}{2}\rho_{bc}^{(n)}-\frac{\gamma}{8} \rho_{bc}^{(n)}, \label{eqbcn}\\
\dot\rho_{cc}^{(n)} & = &-i\Omega(\rho_{bc}^{(n)}-\rho_{cb}^{(n)})-(\Gamma'_{l,\frac{1}{2}}+\Gamma_{r,-\frac{1}{2}}) 
\rho_{cc}^{(n)}+\Gamma_{l,-\frac{1}{2}}\rho_{aa}^{(n)}
+\Gamma'_{r,\frac{1}{2}}\rho_{dd}^{(n-1)}\nonumber \\&-&\frac{\sqrt{\Gamma_{r,\frac{1}{2}}\Gamma_{r,-\frac{1}{2}}}+
\sqrt{\Gamma'_{l,\frac{1}{2}}\Gamma'_{l,-\frac{1}{2}}}}{2}(\rho_{bc}^{(n)}+\rho_{cb}^{(n)}),
\label{eqcn}
\end{eqnarray}
 \begin{eqnarray}
\dot\rho_{dd}^{(n)} & = &-(\Gamma'_{r,\frac{1}{2}}+\Gamma'_{r,-\frac{1}{2}})\rho_{dd}^{(n)}
+\Gamma'_{l,-\frac{1}{2}}\rho_{bb}^{(n)}+\Gamma'_{l,\frac{1}{2}}\rho_{cc}^{(n)}+\sqrt{\Gamma'_{l,\frac{1}{2}}\Gamma'_{l,-\frac{1}{2}}}
(\rho_{bc}^{(n)}+\rho_{cb}^{(n)}),
\label{eqdn} 
\end{eqnarray}
where $\delta=(E_{\frac{1}{2}}-E_{-\frac{1}{2}})/\hbar$ is the difference of the energy levels, 
which are renormalized by the Lamb-shifts. Due to the large bias condition, the other off-diagonal elements, as $\rho^{(n)}_{ad}$, are 
weakly coupled to the differential equations found above and they are not taken in consideration. However, they have their own dynamics, too. 

The left tunneling rates are
\begin{eqnarray}
\label{Gammal}
 \Gamma_{l,s}&=&2 \pi \rho_{l}(E_s) | \omega_{l,s}(E_s) |^2,  
\\
 \Gamma'_{l,s}&=&2 \pi \rho_{l}(E_s+U) | \omega_{l,s}(E_s+U) |^2, 
\nonumber
\end{eqnarray}
and the right tunneling rates are
\begin{eqnarray}
\label{Gammar}
 \Gamma_{r,s}&=&2 \pi \rho_{r}(E_s) | \omega_{r,s}(E_s) |^2, 
 \\
 \Gamma'_{r,s}&=&2 \pi \rho_{r}(E_s+U) | \omega_{r,s}(E_s+U) |^2. 
\nonumber
\end{eqnarray}
where $\rho_{l(r)}$ is the spin up or spin down density of states in the left (right) lead, $\rho_{l(r)}=\rho_{l(r),\frac{1}{2}}=
\rho_{l(r),-\frac{1}{2}}$.\\ 
The energy dependence of the left tunneling amplitudes $\omega_{l,\frac{1}{2}}$ and $\omega_{l,-\frac{1}{2}}$ is a decreasing function, because the 
lowest is the energy level, the highest is the probability to be loaded from the left lead. The energy dependence of the right tunneling 
amplitudes $\omega_{r,\frac{1}{2}}$ and $\omega_{r,-\frac{1}{2}}$ is a increasing function, because the highest energy levels are more 
likely emptied
to the right lead then the lower ones.
Using the relation $E_{\frac{1}{2}}>E_{-\frac{1}{2}}$ we have the following properties for the incoherent tunnelings:
\begin{eqnarray}
\Gamma_{l,-\frac{1}{2}}&>&\Gamma_{l,\frac{1}{2}}>\Gamma'_{l,-\frac{1}{2}}>\Gamma'_{l,\frac{1}{2}}, \nonumber \\ 
\Gamma'_{r,\frac{1}{2}}&>&\Gamma'_{r,-\frac{1}{2}}>\Gamma_{r,\frac{1}{2}}>\Gamma_{r,-\frac{1}{2}}. \nonumber
\end{eqnarray}
We may assume without loss of generality that the probability of filling up the state $\ket{d}=\ket{\uparrow \downarrow}$ from the left lead 
is equal to the probability of emptying the state $\ket{c}=\ket{\downarrow}$ to the right lead, which leads to the assumptions:
\begin{eqnarray}\label{param}
\Gamma_1&=&\Gamma_{l,-\frac{1}{2}}= \Gamma'_{r,\frac{1}{2}},\,\,\,\,\Gamma_2=\Gamma_{l,\frac{1}{2}}=\Gamma'_{r,-\frac{1}{2}}, \\
\Gamma_3&=&\Gamma'_{l,-\frac{1}{2}}=\Gamma_{r,\frac{1}{2}},\,\,\,\, \label{param2} \Gamma_4=\Gamma'_{l,\frac{1}{2}}=\Gamma_{r,-\frac{1}{2}}.
\end{eqnarray}
We remind the reader that we also considered a low temperature case and the large bias condition 
$\mu_l-\mu_r \gg \sum_s E_s + U$ was used. The following calculations will be based 
entirely on the parameters of eqs. \eqref{param}, \eqref{param2}.\\
~\indent
The time evolution of the density matrix is represented in the terms of the number of electrons tunneled through the dot. The convenient 
measurement would be the number of the accumulated electrons, but the number operator $\hat{N}=\sum_{r,s} \hat{a}^\dagger_{r,s} \hat{a}_{r,s}$
has a spectral decomposition, where the different spin states projectors has the same eigenvalue. If this eigenvalue is detected, the detector
does not give an information as to which state belong and therefore we can not determine the state of the dot. Instead of the charge 
measurement we consider the measurement of electric current, which is given by a commutator of $\hat{N}$ with the total Hamiltonian of 
the system, 
\begin{equation}
\hat{I}=i\frac{e}{\hbar}
\big[\hat{H},\hat{N} \big], 
\end{equation}
where $e$ is the elementary charge. Using eq. \eqref{Hamiltonian} we obtain
\begin{equation}
\label{currentop}
\hat{I}= i\,e\, \sum_{r,s} \left(\omega^*_{r,s} \hat{a}^\dagger_{D,s}
\hat{a}_{r,s}-\omega_{r,s}\hat{a}^\dagger_{r,s}\hat{a}_{D,s}\right).
\end{equation}
It follows form eq. \eqref{currentop}, that by measuring the electric current, the projections into different state correspond to different
observed value of the current. This implies that the states of the quantum dot can be determined by monitoring directly the electric 
current.\\    
~\indent    
The model describes the spin-dependent quantum dot, where the magnetization is continuously detected with an averaged 
output gained and at the same time there is a sharp detection of the electric current. The sharp detection gives information about the 
system, and the result will depend on the interaction strength of the continuous detection. 

\section{Results}
\label{Results}
\begin{figure}[t]
\begin{center}
\includegraphics[width=3.1in]{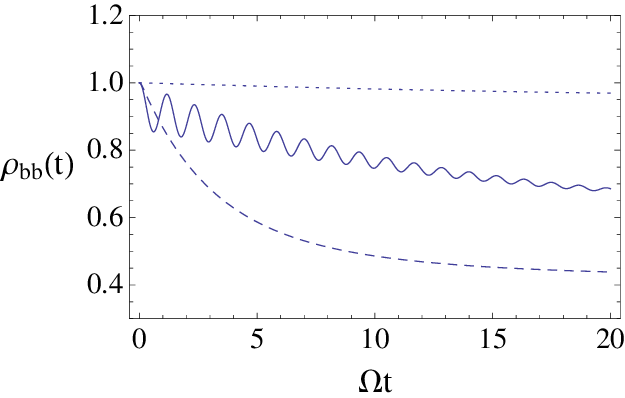}
\caption{\label{fig-2} The occupation of the spin up state as a function of time in the case of \textit{magnetic leads}.  
$\Gamma_4/\Omega=0.04$, $\Gamma_3=0$, $\Gamma_2/\Omega=0.08$, $\Gamma_1=0$ and $\delta/\Omega=5$.
 The curves correspond to different values of the  $\gamma$ parameter: $\gamma=0$ (solid), 
 $\gamma/8=10\Omega$ (dashed), and $\gamma/8=10^3 \Omega$ (dotted).} 
\end{center}
\end{figure}

\begin{figure}[t]
\begin{center}
\includegraphics[width=3.1in]{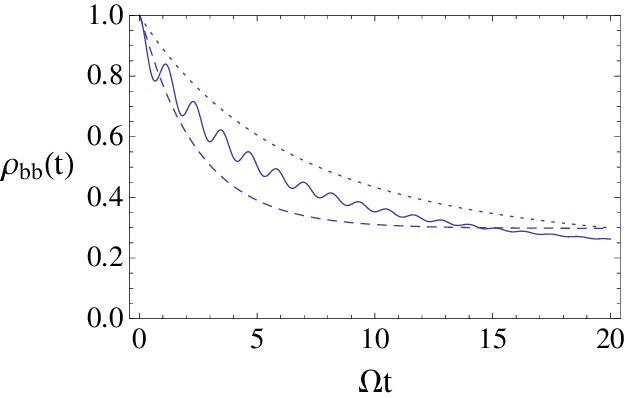}
\caption{\label{fig-1} The occupation of the spin up state as a function of time in the case of \textit{normal leads}.
 $\Gamma_4/\Omega=0.04$, $\Gamma_3/\Omega=0.06$, $\Gamma_2/\Omega=0.08$, 
$\Gamma_1/\Omega=0.1$ and $\delta/\Omega=5$.
 The curves correspond to different values of the  $\gamma$ parameter: $\gamma=0$ (solid), 
 $\gamma/8=10\Omega$ (dashed), and $\gamma/8=10^3\Omega$ (dotted).} 
\end{center}
\end{figure}

\begin{figure}[t]
\begin{center}
\includegraphics[width=3.1in]{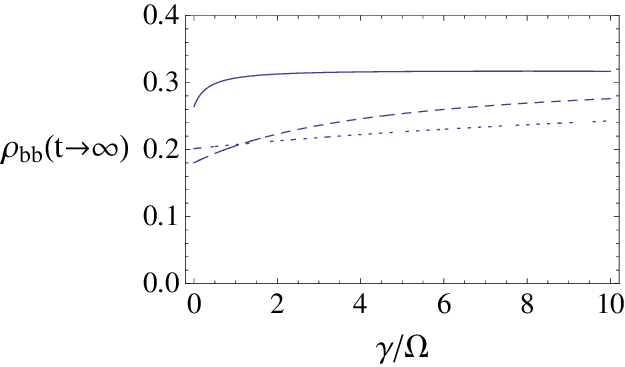}
\includegraphics[width=3.1in]{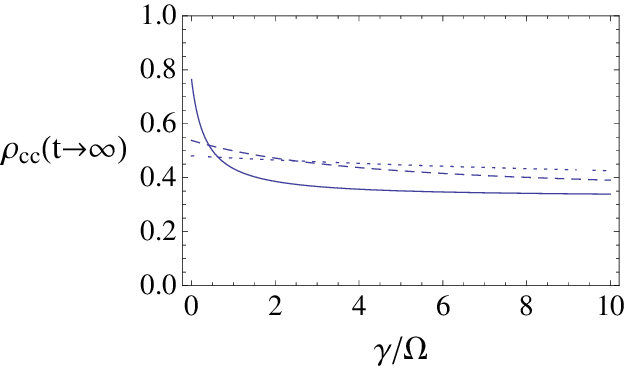}
\caption{\label{fig0} Probability of the $|b \,\rangle$(spin up) and $|c\, \rangle$(spin down) steady states as a function of the 
rescaled damping parameter. 
$\Gamma_1/\Omega=0.1$, $\Gamma_2/\Omega=0.08$, $\Gamma_3/\Omega=0.06$ and $\Gamma_4/\Omega=0.04$. 
The curves correspond to different values of the rescaled energy difference: $\delta/\Omega=1$ (solid), $\delta/\Omega=5$ (dashed), 
and $\delta/\Omega=10$ (dotted).} 
\end{center}
\end{figure}

We are going to analyze two cases to show the presence of measurement back-action and QZE. We remind the reader that the unsharply 
detected operator, the magnetization, is diagonal in the Hilbert space of the dot and this is the 
reason why the damping mechanism has effect only on the internal coherent motion. The theory of the continuous quantum measurement allows
the discussion of more general operators, which may introduce a complex damping mechanism, although they have to be subject of 
possible experimental realizations. \\
~\indent 
QZE is where the repeated observations of the system slow down transitions between quantum states. As a result of a continuous 
observation the system cannot leave its initial state \cite{misra}. The original paper formulates the problem in very general way, but
only proves for projective measurements, which prevails negligibly small time evolution during the time $t/n$ when the number of repetitions
$n \to \infty$. The theory of continuous quantum measurement 
is built up from sequence of unsharp measurements, such that each measurement 
is increasingly weak by the increase of the repetition $n$. This construction \cite{BLP1,BLP2} allows QZE only for 
measurement strength $\gamma \to \infty$.  In order to investigate the effect, let us consider the case that 
our system is initially in the spin up state.\\
~\indent
In the first step we study our model in the presence of magnetic leads. We define the left lead with spin up, and the right lead with 
spin down states: $\Gamma_1=\Gamma_3=0$. Before any further discussion we need a reminder  that the direction of the current was fixed, 
and is flowing through the dot from left to right. Here, the electrons are coming from the left and fill up the $|b \,\rangle$ state 
then they hop to the $|c \,\rangle$ state, from where they leave the dot to the right lead. When the measurement-induced damping 
parameter $\gamma/8$ is smaller than the spin flip frequency $\Omega$, the system oscillations are maintained. If $\gamma$ is increased, the 
coherent oscillations die out, see Figs. \ref{fig-2}. We find that for small $t$ the rate of transition form the spin up to the spin down 
state slows down with the increase of $\gamma$. If $\gamma$ tends to infinite then the off-diagonal density matrix elements, 
eq. \eqref{eqbcn} and its complex conjugate, do not 
contribute to the dynamics and the system remains frozen in its initial state, the state $|b \,\rangle$. This implies that continuous 
measurements would localize the system and our equations reproduce the QZE. If we consider the spin down state as the initial state, 
then the system will not remain there. As $\gamma$ is increased the transition from the spin down state to the the spin up state is 
slowed down, although the system will be localized in the state $|b \,\rangle$ at $t \to \infty$. 
%%%%%%%%%%%%%%%%%%%%%%%%%%%%%%%%%%%%%%%%%%%%%%%%%%%%%%%%%%%%%%%%%%%%%%%%%%%%%%%
%%%%%%%%%%%%%%%%%%%%%%%%%%%%%%%%%%%%%%%%%%%%%%%%%%%%%%%%%%%%%%%%%%%%%%%%%%%%%%%%%%
This is a direct consequence of the conditions imposed, which result that the state $|b \,\rangle$ is the preferred steady-state 
due to the combination of right lead magnetization and the direction of tunneling.\\
~\indent
While for small $t$ the transition is slowed down, after some critical time $t>t_0$ an enhanced decay can be seen (Figs. \ref{fig-2}) 
for $\gamma$ values comparable to the energy level displacement $\delta$. This enhanced decay results faster transition between the 
spin up and the spin down states than in the unmeasured system. Nevertheless the effect is just the opposite of the QZE. These behaviours 
are also reflected in the steady-state. 
If the enhanced decay occur then the probability of the state $|b \,\rangle$ is smaller than in the case of the unmeasured system. \\
~\indent
In spite of great progress made in microfabrication techniques, the construction of magnetic leads to a tiny quantum dot is still a 
challenge and we also study the considered system with normal leads. The localization into the state $|b \,\rangle$ 
as in the magnetic lead model is impossible, because both the spin up and spin down states can be filled from left and emptied to right. 
Examining the results in Figs. \ref{fig-1} the effect of suppressed transition can be found for small values of $t$. Here, the system 
cannot be frozen in its initial 
state as a consequence of the normal leads. When $\gamma$ is comparable with $\delta$ then the effect of the enhanced decay 
characterize the time evolution of the state $|b \,\rangle$.
The steady-state behaviour shows a different aspect than in the case of magnetic leads. If the strength of the measurement increases then the 
probability of the state $|b \,\rangle$ is increasing. In the case of the 
unmeasured system, the probability of the steady state $|b \,\rangle$ may decrease as a function of $\delta$, but for large values of $\gamma$
is an increasing function of $\delta$. This means, when the suppression of the coherent oscillations is weak and the energy difference $\delta$
is large
enough, then the probability of filling up the state $|b \,\rangle$ is less likely. As we expect, the probability of the steady state 
$|c \,\rangle$, the lower energy level, has the opposite character as a function of $\gamma$. These behaviours are shown in 
Fig. \ref{fig0}. Here, exists a mixed state, namely $\lim_{\gamma \to \infty}\hat{\rho}(t \to \infty)$, which provides the QZE. If this state
can be prepared as an initial condition, then for the case of no magnetization measurement the system evolves into another mixed state, but
for infinite accuracy measurement $\Delta M \to 0$ (eq. \eqref{gamma}), $\gamma \to \infty$, the system remains frozen in its initial condition. \\
~\indent   
Next we analyze the electric current flowing through the system. This quantity is the only measurement 
result retained and we study it, to show the presence of measurement back-action. Using eqs. \eqref{Psi}, \eqref{currentop} we obtain 
the average electric current \cite{averageinner}:
\begin{eqnarray}\label{current}
  &&\bra{\Psi(t)}\hat{I}\ket{\Psi(t)}=e\Big(\Gamma_3 \sum_n \rho^{(n)}_{bb}(t)+\Gamma_4 \sum_n \rho^{(n)}_{cc}(t)  \nonumber \\
&&+2 \sqrt{\Gamma_3 \Gamma_4}\sum_n \mathrm{Re}[\rho^{(n)}_{bc}(t)] +(\Gamma_1+\Gamma_2)\sum_n \rho^{(n)}_{dd}(t) \Big). \nonumber \\
\end{eqnarray}
%%%%%%%%%%%%%%%%%%%%%%%%%%%%%%%%%%%%%%%%%%%%%%%%%%%%%%%%%%%%%%%%%%%%%%%%%%%%%%%
%%%%%%%%%%%%%%%%%%%%%%%%%%%%%%%%%%%%%%%%%%%%%%%%%%%%%%%%%%%%%%%%%%%%%%%%%%%%%%%%%%
%%%%%%%%%%%%%%%%%%%%%%%%%%%%%%%%%%%%%%%%%%%%%%%%%%%%%%%%%%%%%%%%%%%%%%%%%%%%%%%
%%%%%%%%%%%%%%%%%%%%%%%%%%%%%%%%%%%%%%%%%%%%%%%%%%%%%%%%%%%%%%%%%%%%%%%%%%%%%%%%%%
%%%%%%%%%%%%%%%%%%%%%%%%%%%%%%%%%%%%%%%%%%%%%%%%%%%%%%%%%%%%%%%%%%%%%%%%%%%%%%%
\begin{figure}[t]
\begin{center}
\includegraphics[width=3.1in]{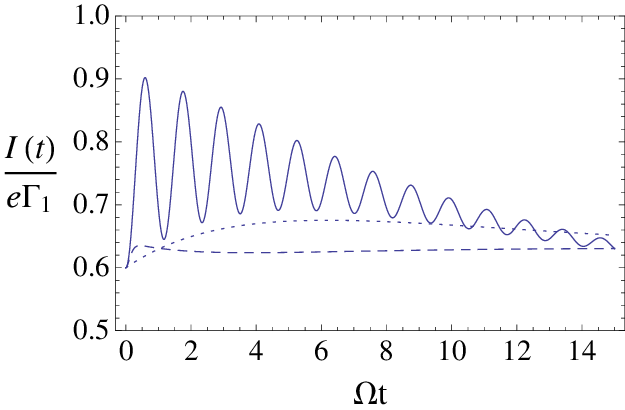}
\caption{\label{currenttime} Electron current through the dot as a function of time.  $\Gamma_4/\Omega=0.04$, $\Gamma_3/\Omega=0.06$, 
$\Gamma_2/\Omega=0.08$, $\Gamma_1/\Omega=0.1$ and $\delta/\Omega=5$.
 The curves correspond to different values of the  $\gamma$ parameter: $\gamma=0$ (solid), 
 $\gamma/8=10\Omega$ (dashed), and $\gamma/8=10^3 \Omega$ (dotted).} 
\end{center}
\end{figure}
%%%%%%%%%%%%%%%%%%%%%%%%%%%%%%%%%%%%%%%%%%%%%%%%%%%%%%%%%%%%%%%%%%%%%%%%%%%%%%%%%%
In Fig. \ref{currenttime} we find the suppression effect induced by the continuous measurement. In the case of no measurement the 
current is increasing and oscillating and then is followed by a decrease and relaxation to the steady-state. If the enhanced 
decay occur in the spin up state then the current shows a slow increase and a quick relaxation to its stationary value, which is 
bigger than the value found for the unmeasured case. If the measurement is extremely strong then the current is increasing slowly 
compared to the other cases, the relaxation is longer and the asymptotic value is the highest. This means that the suppression of
transitions can be determined
directly after a short time evolution of the system or indirectly by examining the relaxation mechanism. The electric current in the 
magnetic leads case is zero, 
when QZE occurs. \\  
~\indent
In a future experimental setup where this double detection setup will be applied, changing the conditions of the  unsharp 
detection and analyzing the sharp detection output, the scope of the continuous quantum measurement theory may be demonstrated. 
The substrate induced decoherence was treated, so a possible charge current change will only be the result of a measurement back-action.  
\\ 

\section{The accuracy of the measurement}
\label{accuracy}
~\indent
We analyzed the detection of measurement back-action by the average electric current, which can be determined from ensemble
measurements. The latter involves the problem of accuracy. If the current can be detected directly, via its magnetic field \cite{Levitov}, 
then the states of the dot can be monitored with any accuracy.\\
~\indent
If such a measurement cannot be performed, one can make an indirect measurement. Obtaining the charge counting statistics in the right 
lead is the most plausible in this systems and from the statistics can be deduced the average current. Recently, there was another suggestion, 
where the variation of the right lead charge was discussed \cite{Gur2008}. In this cases the standard deviation $\sigma_{\hat{I}}$
of the electric current from its mean
can be best understood from the uncertainty principle. The uncertainty principle gives a lower bound for the standard deviation. The 
upper bound of the standard deviation is the square root of the second moment. The standard deviation could be any number between 
these bounds. However, we focus on the lower bound, because this quantity tells us when the measurement of the current could be 
less uncertain.\\
~\indent
For the observables $\hat{I}, \hat{N}$ and the pure state $\ket{\Psi}$ the following 
inequality holds:
\begin{equation}
\label{ineq}
 \sigma^2_{\hat{I}}\sigma^2_{\hat{N}}\geq \frac{1}{4} \Big|\bra{\Psi} \left [\hat{I},\hat{N}\right] \ket{\Psi}\Big|^2,
\end{equation}
where (see \cite{averageinner}) 
\begin{eqnarray}
\sigma^2_{\hat{I}}&=&\bra{\Psi}\hat{I}^2 \ket{\Psi}-\bra{\Psi}\hat{I} \ket{\Psi}^2,\\
\sigma^2_{\hat{N}}&=&\bra{\Psi}\hat{N}^2 \ket{\Psi}-\bra{\Psi}\hat{N} \ket{\Psi}^2.
\end{eqnarray}  
We calculate the right hand side of the inequality by using the eqs. \eqref{Psi},\eqref{currentop}:
\begin{eqnarray}
&&\left [\hat{I},\hat{N}\right]=i\,e\, \sum_{r,s} \left(\omega^*_{r,s} \hat{a}^\dagger_{D,s}
\hat{a}_{r,s}+\omega_{r,s}\hat{a}^\dagger_{r,s}\hat{a}_{D,s}\right), \\
&&\bra{\Psi} \left [\hat{I},\hat{N}\right] \ket{\Psi}=-2i\,e\sqrt{\Gamma_3 \Gamma_4}\sum_n \mathrm{Im}[\rho^{(n)}_{bc}(t)] \nonumber \\
&&=-2i\,e\sqrt{\Gamma_3 \Gamma_4}\mathrm{Im}[\rho_{bc}(t)]. 
\end{eqnarray}
In order to evaluate the standard deviation of the charge number, we have to rewrite the 
related averages in the following way
\begin{eqnarray}
\bra{\Psi}\hat{N} \ket{\Psi}&=&\overline{n}_{aa}+\overline{n}_{bb}+\overline{n}_{cc}+\overline{n}_{dd}, \\
\bra{\Psi}\hat{N}^2 \ket{\Psi}&=&\overline{n^2}_{aa}+\overline{n^2}_{bb}+\overline{n^2}_{cc}+\overline{n^2}_{dd}.
\end{eqnarray}
These new averages are defined as $\overline{n}_{ij}=\sum_n n \rho^{(n)}_{ij}$ and $\overline{n^2}_{ij}=\sum_n n^2 \rho^{(n)}_{ij}$.
Multiply eqs. \eqref{eqan}, \eqref{eqbn}, \eqref{eqbcn}, \eqref{eqcn}, \eqref{eqdn} by $n$ or $n^2$ and sum over $n$ one finds
coupled differential equations
\begin{eqnarray}
 \dot\rho_{aa}&=& -(\Gamma_1+\Gamma_2)\rho_{aa}
+\Gamma_3\rho_{bb}+\Gamma_4\rho_{cc} \nonumber \\
&+&\sqrt{\Gamma_3\Gamma_4}
(\rho_{bc}+\rho_{cb}),\\
\dot{\overline{n}}_{aa}&=& -(\Gamma_1+\Gamma_2)\overline{n}_{aa}
+\Gamma_3\overline{n}_{bb}+\Gamma_4\overline{n}_{cc} \nonumber \\
&+&\sqrt{\Gamma_3\Gamma_4}
(\overline{n}_{bc}+\overline{n}_{cb})+\Gamma_3\rho_{bb}+\Gamma_4\rho_{cc} \nonumber \\
&+&\sqrt{\Gamma_3\Gamma_4}
(\rho_{bc}+\rho_{cb}),
\end{eqnarray}
\begin{eqnarray}
 \dot{\overline{n^2}}_{aa}&=&(\Gamma_1+\Gamma_2)\overline{n^2}_{aa}
+\Gamma_3\overline{n^2}_{bb}+\Gamma_4\overline{n^2}_{cc} \nonumber \\
&+&\sqrt{\Gamma_3\Gamma_4}
(\overline{n^2}_{bc}+\overline{n^2}_{cb})+2\Gamma_3\overline{n}_{bb}+2\Gamma_4\overline{n}_{cc} \nonumber \\
&+&2\sqrt{\Gamma_3\Gamma_4}
(\overline{n}_{bc}+\overline{n}_{cb})+\Gamma_3\rho_{bb}+\Gamma_4\rho_{cc} \nonumber \\
&+&\sqrt{\Gamma_3\Gamma_4}
(\rho_{bc}+\rho_{cb}), \\
\ldots \nonumber
\end{eqnarray}
where $\rho_{ij}=\sum_n \rho^{(n)}_{ij}$.
We rewrite eq. \eqref{ineq} as
\begin{equation}
\label{44}
\sigma_{\hat{I}}\geq  \frac{e\sqrt{\Gamma_3 \Gamma_4}\Big|\mathrm{Im}[\rho_{bc}(t)]\Big|}{\sqrt{\sum_i\overline{n^2}_{ii}(t)-
\Big(\sum_i \overline{n}_{ii}(t)\Big)^2}},\,\, i \in \{a,b,c,d\}, 
\end{equation}
where the right hand side of the inequality is given by the coupled differential equations found above.\\
%%%%%%%%%%%%%%%%%%%%%%%%%%%%%%%%%%%%%%%%%%%%%%%%%%%%%%%%%%%%%%%%%%%%%%%%%%%%%%%
\begin{figure}[t]
\begin{center}
\includegraphics[width=3.1in]{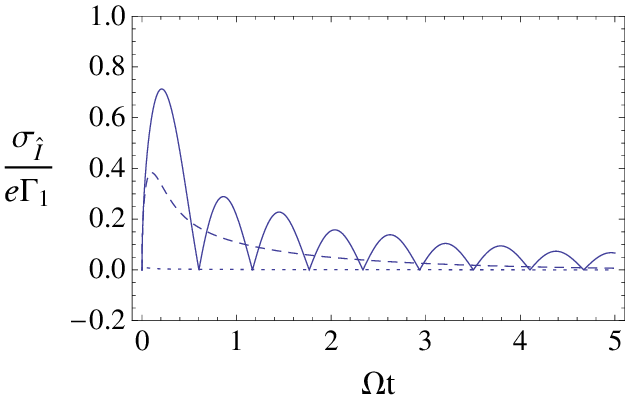}
\caption{\label{uncertainplot} The lower bounds of the standard deviation (electric current) as a function of time.  
$\Gamma_4/\Omega=0.04$, $\Gamma_3/\Omega=0.06$, 
$\Gamma_2/\Omega=0.08$, $\Gamma_1/\Omega=0.1$ and $\delta/\Omega=5$.
 The curves correspond to different values of the  $\gamma$ parameter: $\gamma=0$ (solid), 
 $\gamma/8=10\Omega$ (dashed), and $\gamma/8=10^3 \Omega$ (dotted).} 
\end{center}
\end{figure}
%%%%%%%%%%%%%%%%%%%%%%%%%%%%%%%%%%%%%%%%%%%%%%%%%%%%%%%%%%%%%%%%%%%%%%%%%%%%%%%%%%
~\indent
We consider that the system is initially in the spin up state. Figs. \ref{uncertainplot} shows that the standard deviation of the electric
current is decreased in time. However, the effect in interest, induced by the continuous measurement, takes place for short times.
Figs. \ref{uncertainplot} also shows that in the case of $\gamma/8 \gg \Omega$, the lowest possible value of 
$\sigma_{\hat{I}}$ 
is small for short times. This is not
surprising(see eq. \eqref{44}) since the large decoherence generated by the magnetization detector, reduces the off-diagonal elements to zero.  
It follows from our
analysis that only highly accurate measurement cases (see eq. \eqref{gamma}) could have small standard deviations. Having two different 
setup with $\gamma$ and $\gamma'$ 
parameters and 
implementing the $\gamma,\gamma' \gg \Omega$ condition, the rate of the suppression can also be compared.   

\section{Conclusions}
\label{Conclusions}
In this paper we considered a simple spin-dependent quantum dot and we studied the mechanism of continuous magnetization measurement
by a detected electric current.  Starting with the 
many-particle wave function in the occupation number representation and adding the dynamics of the continuous measurement, we have found 
the equations of motion for the system. Our derivation contains an important
assumption, namely that the energy states of the system are deep inside the bias $\mu_l-\mu_r$. If this condition is not fulfilled, then
more off-diagonal elements may appear in the equation of motion, and new couplings between different density matrix elements may occur. 
The continuous measurement of the magnetization induce a damping rate in the equations for the off-diagonal density-matrix 
elements. The analysis revealed that the states of the dot can be determined only by a electric current measurement, 
due to the spectral decomposition of the current operator.\\
~\indent  
The back-action of the continuous detection damps the oscillations 
between the different spin states. We investigated the case of magnetic leads, where an enhanced decay takes place when the 
damping parameter $\gamma$ is comparable with the energy level displacement $\delta$.
The spin up state is the preferred steady-state as the consequence of our choice of lead magnetization and the tunneling direction. 
If this state is the initial state of the system, then for very strong measurements the system remains frozen there, showing the presence of 
quantum Zeno effect. 
We found that the damping mechanism in the case of normal leads is also suppressing the coherent motion and inducing an enhanced decay, too. 
Here, the spin up state is no more a preferred steady-state and for strong measurements the system tends to a 
particular mixed state. The behaviour of the enhanced decay and the strong damping is reflected also in the steady-state of the 
system. When the enhanced decay occurs the steady-state values are smaller (state $|b \,\rangle$) or higher 
(state $|c \,\rangle$) than in the case of the unmeasured system. If $\gamma/8<\delta$, then there is a faster increase in the 
steady-state $|b \,\rangle$ as a function of $\gamma/\Omega$. \\
~\indent 
We studied the electric current as the only quantity retained by the detector. The increase rate of the electric current for 
short times depends
inversely proportional on the continuous measurement parameter $\gamma$. If the parameter $\gamma$ increase, the rate of the 
current is decreasing. The current shows a quick relaxation to its 
stationary value, when the enhanced decay occurs in the spin states. In the case of strong damping, 
a long relaxation characterize the time evolution of the electric current. The results also show that the coherent oscillations die out
for $\gamma/\Omega \gg 1$. When quantum Zeno effect occurs in the case of magnetic leads, then the electric current is zero. \\
~\indent
In a real experiment the mean value of the electric current is deduced form the charge counting statistics. This shows a necessity of 
studying the standard deviation of the electric current from its mean. We derived a lower bound for the deviation, 
using the Heisenberg uncertainty principle. We found that only for highly accurate magnetization measurements is possible to
analyze measurement back-action and quantum Zeno effect by the detection of the electric current. \\
~\indent
Experimentally, the analyzed model can be implemented by semiconductor spin-light emitting diode structures
containing single layers of InAs/GaAs self-assembled quantum dots \cite{Experiment}. Application of a oblique magnetic field 
($\overrightarrow{B}(B_x, 0, B_z)$
in the model) results the emission of a circular polarized light \cite{Dykonov}. The emitted light is used for measuring the polarization of 
the injected spin or the spin dynamics in the quantum dot \cite{Motsnyi}. The photon detector in these experiments is the apparatus, which
we modeled as the unsharp detector in the theory presented. The applied oblique magnetic field is 40-50 mT, which results a Larmor frequency $\Omega=
g \mu_B B /\hbar \sim 10^8 s^{-1}$, where $g$ is the effective Land\'e factor and $\mu_B$ is the Bohr magneton. The characteristic time
of a tunneling process is $\tau \sim \mu s - ms$ and a typical tunneling rate $\Gamma \sim 10^3 - 10^6 s^{-1}$. The time scale of the 
inverse measurement parameter $1/\gamma$ and of the photon emission process can be taken to be the same. The emission time $\tau \sim ns$ 
and then $\gamma \sim 10^9 s^{-1}$. $\gamma$ depends on the accuracy and the bandwidth of the photon detector, so $10^9 s^{-1}$ should be 
considered as the lower bound for this parameter. These parameters are adjustable, so the requirements of our analysis might be achieved by a 
real experimental setup. There are other decoherence sources in these devices, as hyperfine interactions with unpolarized nuclei,
the temperature dependent phonon absorption and emission. These decoherences are competing with the measurement induced decoherence. 
However, their decoherence rate cannot be manipulated at will, so the effects induced by the adjustable $\gamma$ should be filtered out 
by a series of measurements.\\
~\indent      
The model offers some insight into the measurement induced back-action and studies the way the QZE appears in a complex system. 
An experimental study has the potential to analyze the scope of the continuous quantum measurement theory in mesoscopic solid state 
physics. Tight control of systems parameters may strongly enhance the possibilities of observing the measurement-induced back-action.\\ 

The author would like
to thank S. A. Gurvitz for discussions and encouragement.
J. Z. B. has profited also from helpful suggestions made by T. Geszti, M. J\"a\"askel\"ainen, and U. Z\"ulicke. This work was supported 
by a postdoctoral fellowship grant from the Massey University Research Fund. Additional funding through BMBF project QK{\_}QuOReP is 
gratefully acknowledged.

\appendix
 \section{General measurements}
 \label{GM}
Usually the von Neumann measurement that is discussed, when the system is projected onto one
of the possible eigenstates of a given observable. Considering
these eigenstates as $\{|n\rangle :
 n=1,\ldots,d\}$ 
and the state of the system $|\psi\rangle = \sum_n a_n |n\rangle$, then the system is randomly projected onto $|n\rangle$ 
with the probability $p_n=|a_n|^2$.  The properties of the projectors are:
\begin{eqnarray}
 \sum_n\hat{P}_n&=&\hat{I},\\
\hat{P}_n\hat{P}_n&=&\delta_{n,m}\hat{P}_n.
\end{eqnarray}
These sharp measurements can be discussed in two ways. {\em Selective\/} case is when the system is randomly projected onto an eigenstate
\begin{eqnarray}
\hat{\rho} &\rightarrow &\hat{\rho}_n=
\frac{\hat{P}_n \hat{\rho} \hat{P}_n}{\mathrm{Tr} [\hat{P}_n \hat{\rho} \hat{P}_n] }, \\ 
\hat{A} &\rightarrow & A_n,
\end{eqnarray}
where the possibility to detect eigenvalue $A_n$ is $p_n=\mathrm{Tr} [\hat{P}_n \hat{\rho} \hat{P}_n]$.\\
{\em Nonselective\/} description represent a measurement scenario, where our record of the result $A_n$  was lost.
We have a quantum system in the state $\hat{\rho}_n$ with probability $p_n$, but we no longer know the actual value of $A_n$. 
The state of such a quantum system is the mixture of the states $\hat{\rho}_n$ with probabilities $p_n$. Due to the rules of 
probability theory, the following results are obtained:
\begin{eqnarray}
\label{eq1}
 \hat{\rho} &\rightarrow &\hat{\rho}_n=\sum_n p_n \hat{\rho}_n=
\sum_n \hat{P}_n \hat{\rho} \hat{P}_n, \\ 
\hat{A} &\rightarrow & \sum_n p_n A_n=\langle \hat{A}\rangle.
\end{eqnarray}
 A von Neumann measurement
provides complete information and the system is always projected into an eigenstate. 
However, there exist many measurements, which are unable to detect sharply these eigenvalues. If the detector plus 
the system is under a projective measurement scenario, then the larger system will act on the system in ways that cannot 
be described by projective measurement on the system alone. For these purposes we need to consider a general class of measurements, which
result reduced quantity of information about an observable.\\

These {\em unsharp\/}  measurements can be described by generalizing the set of projectors. The construction can be done in two 
ways: the cardinality of the sets are countable or infinite.
Suppose we pick a set of $N$ 
operators $\hat{\Pi}_n$, the
restrictions are: $\sum_{n=1}^{N} \hat{\Pi}_n
= \hat{I}$, where $\hat{I}$ is the identity operator and $\hat{\Pi}_n$ to be hermitian positive semidefinite. 
A hermitian positive semidefinite operator can always be written as, $\hat{\Pi}_n = \hat{M}^\dagger_n \hat{M}_n $,
for some operator $\hat{M}_n$. If the positivity of $\hat{M}_n$ is not required, the square root of $\hat{\Pi}_n$ can give infinite 
solutions, which means that there are infinite different experimental apparatuses that gives the same probabilities for the outcomes.\\
The selective description is:
\begin{equation}
\hat{\rho}\rightarrow \hat{\rho}_n=
\frac{\hat{M}_n \hat{\rho} \hat{M}^\dagger_n}{\mathrm{Tr} [\hat{M}_n \hat{\rho} \hat{M}^\dagger_n] }, 
\end{equation} 
with
\begin{eqnarray}
 p_n &=& \mathrm{Tr}[\hat{M}_n \hat{\rho} \hat{M}^\dagger_n],\\
\sum_{n=1}^{N} p_n&=&1,
\end{eqnarray}
giving the probability of obtaining the $n$th outcome. \\
The nonselective description is:
\begin{equation}\label{eq2}
\hat{\rho}\rightarrow \sum_{n=1}^{N} p_n \hat{\rho}_n=
\sum_{n=1}^{N} \hat{M}_n \hat{\rho} \hat{M}^\dagger_n. 
\end{equation} 
The set can be countable infinite, here we have $N=\infty$, but also uncountable infinite is possible where the sum will be 
replaced by an integral and the normalization is:
\begin{equation}
 \int p(x) dx =1.
\end{equation}
These generalized measurements are called as POVM's (positive operator-valued measure). 
The mappings of the density matrices in eqs. \eqref{eq1}, \eqref{eq2} has a specific form, called Kraus-form \cite{Kraus}.

 \section{Continuous quantum measurement of an observable}
 \label{CQM}
A continuous measurement is referring to a scenario, where the measurement data is
continually extracted from a system. In order to construct a measurement like this, we consider the partition
of the time line into a sequence of intervals of length $\Delta t$,
and consider an unsharp measurement in each interval. In this construction, there is a mathematical requirement for obtaining 
a dynamical equations, namely the strength of each measurement
must depend on the length of the time interval. This construction circumvent the problem of instantaneous effect of a standard 
quantum measurement.

We now divide time into intervals of length $\Delta t$. In each
time interval, we will make a measurement described by the
operators
\begin{equation}
\hat{\Pi}(x)=\frac{1}{\sqrt{2\pi\sigma^2}}
\mathrm{exp}\left[-\frac{(x-\hat{A})^2}{2\sigma^2}\right],\,\, -\infty\leqslant x \leqslant\infty.
  \label{povm}
\end{equation}
Each operator $\hat{\Pi}(x)$ a Gaussian-weighted sum of projectors  
onto the eigenstates of $\hat{A}$ and the probability  
\begin{equation}
 p(x)=\mathrm{Tr}\left(\hat{\Pi}(x)\hat{\rho}\right),\,\, \int p(x) 
dx=1.
\end{equation}
The observable $\hat{A}$ can be \textit{any} kind of Hermitian operator.
The following equation is also true:
\begin{equation}
\label{expectation}
\langle x \rangle=\int x p(x) dx=\langle \hat{A} \rangle. 
\end{equation}
The strength of the continuous measurement is $\gamma$ defined by
\begin{equation}
 \gamma=\frac{1}{\Delta t \,\sigma^2}.
\end{equation}
The \textit{selective} description of such a measurement (eq. \eqref{povm}) is derived by introducing $x$ as a stochastic quantity. 
Main reason for this is to keep the random nature of the measurement
and due to eq. \eqref{expectation} we can write:
\begin{equation}
x(t+\Delta t)-x(t) = \langle \hat{A} \rangle\, \Delta t + \frac{W_{t+\Delta t}-W_t}{\sqrt{\gamma}},
\end{equation}
where $W$ is the Wiener process, a Gaussian random variable. The measurement result gives the
average value $\hat{A}$ in the time interval $\Delta t$ but also a random value due to the form of $\hat{\Pi}(x)$.
 
Now, we make a sequence form the aforementioned measurements and take the limit $\Delta t \rightarrow 0$. 
As the limit is taken, then $\hat{\Pi}(\Delta t \to 0)\to 0$ in eq. \eqref{povm}.
To derive a dynamical equation, it is enough to consider the change induced by operator $\hat{\Pi}(x)$, since the operator is the function of 
$\Delta t$ and does not change during the time interval $t$. The measurement induced change is derived to the first order in $\Delta t$. 
The state after the measurement yields
\begin{equation}
 | \Psi(t+\Delta t) \rangle=\frac{1}{\sqrt{p(x)}}
\hat{\Pi}(x)^{1/2} | \Psi(t) \rangle.
\end{equation}
Expanding $\hat{\Pi}(x)^{1/2}$ 
to first order in $\Delta t$, leads to the relation
\begin{equation}
\mid \Psi(t+\Delta t) \rangle \varpropto \left[1-\frac{\gamma}{8}\left(\langle \hat{A}\rangle-\hat{A}
\right)^2 \Delta t-\frac{\sqrt{\gamma}}{2}\left(\langle \hat{A}\rangle-\hat{A}
\right)\Delta W\right] \mid \Psi(t) \rangle, 
\end{equation}
where we used that in the limit of $\Delta t
\rightarrow 0$, $(\Delta W)^2 \rightarrow \Delta t$. These equation does not preserve the norm of the wave function, 
so we have to calculate $p(x)$ in the first order expansion in $\Delta t$, including the stochastic calculation rule induced above.  
Now we take the limit $\Delta t\rightarrow 0$, setting $\Delta t = dt$, 
$\Delta W = dW$ and $(\Delta W)^2 = dt$ we get
\begin{eqnarray}
\label{SS}
&&|\Psi(t+\Delta t) \rangle=| \Psi(t) \rangle+d | \Psi \rangle,\\
&&d | \Psi \rangle=\left[-\frac{\gamma}{8}\left(\langle \hat{A}\rangle-\hat{A}
\right)^2dt-\frac{\sqrt{\gamma}}{2}\left(\langle \hat{A}\rangle-\hat{A}
\right)dW\right] | \Psi \rangle. \nonumber 
\end{eqnarray}
This is the equation which describes the
evolution of the state of a system in a time interval $dt$ and it is called stochastic Schr\"odinger equation. 
The state $|\psi \rangle$ evolves randomly. An equation for the density operator
$\hat{\rho}$  can be derived, too. Using the same stochastic calculation rules, and defining $\hat{\rho}(t+dt) =
\hat{\rho}(t) + d\hat{\rho}$, we have
\begin{equation}
\label{SM}
 d\hat{\rho}=-\frac{\gamma}{8} \left[\hat{A},\left[\hat{A},\hat{\rho}\right]\right] dt+
\frac{\sqrt{\gamma}}{2}\left(\hat{A}\hat{\rho}+\hat{\rho}\hat{A}-2 \langle \hat{A}\rangle \hat{\rho} \right) dW,
\end{equation}
a stochastic master equation.\\

The \textit{nonselective} description is when the observer makes the continuous measurement, but throws away
the information regarding the measurement results, i.e detecting the average of an observable instead of the eigenvalues, so
the observer must average over the different possible results.
Due to the construction of the model the quantities $\rho$ and $dW$ are statistically independent, $\ll\rho\, dW\gg =
0$ (average over all possible realizations and over the Hilbert space states). Thereby, we can set to zero all terms
proportional to $dW$ in eq. \eqref{SM}, which yields
\begin{equation}
\label{decoherence}
 \frac{d\hat{\rho}}{dt}=-\frac{\gamma}{8}\left[ \hat{A},\left[\hat{A},\hat{\rho}\right]\right].
\end{equation}
The double-commutator describes the decoherence caused by the continuous quantum measurement.\\
Eq. \eqref{decoherence} also can be achieved without introducing the random variables. For a time interval $\Delta t$ the 
nonselective evolution of the density matrix is
\begin{equation}
 \hat{\rho}(t+\Delta t)= \int \hat{\Pi}(x)^{1/2} \hat{\rho}(t) 
\hat{\Pi}(x)^{1/2} dx.
\end{equation}
Expanding this equation into series up to the leading term $\Delta t$ and calculating the integral, the following result can be obtained
\begin{equation}
 \hat{\rho}(t+\Delta t)=\hat{\rho}(t)-\frac{\gamma}{8}\Big(\hat{A}^2 \hat{\rho}(t)-
2 \hat{A} \hat{\rho}(t) \hat{A} + \hat{\rho}(t) \hat{A}^2 \Big) \Delta t.
\end{equation}
Taking the limit $\Delta t \rightarrow 0$ we arrive at the same equation as eq. \eqref{decoherence}.\\

Under unitary evolution, the following transformation is added to eq. \eqref{SS}
\begin{equation}
  | \psi \rangle \rightarrow | \psi \rangle + d|\psi \rangle =
    \left(1-i\frac{\hat{H}}{\hbar}\,dt\right)| \psi \rangle,
\end{equation}
where $\hat{H}$ is the Hamiltonian. For the eqs. \eqref{SM},\eqref{decoherence} the transformation is:
\begin{equation}
    \hat{\rho} + d\hat{\rho} = \hat{\rho}-\frac{i}{\hbar}[\hat{H},\hat{\rho}]\,dt.
\end{equation}
If we want to treat  non-unitary dynamics then the infinitesimal transformation $\mathcal{L} \hat{\rho} \,dt$  can be added only to 
eqs. \eqref{SM} and \eqref{decoherence}. We have to remark also that the form of the equation of motion, namely the appearance of the double
commutator, is the consequence of the Gaussian form of $\hat{\Pi}(x)$.


\begin{thebibliography}{00}
\bibitem{nuclearmagnetic} The effects of the ever-present environmental nuclear magnetic field can be interpreted as an unsharp measurement, 
however this cannot be manipulated at will.
\bibitem{averageinner} In the present formalism it takes some extra care to calculate $\langle .\rangle$. 
This average is not only the quantum mechanical average of an operator, but an average over the random variable, too. Here, the
random variable is the standard Wiener process. The need of a double average is generated by the $\bra{\Psi}\Psi \rangle=1$ condition.
\end{thebibliography}
\end{document}